\documentclass[submission,copyright,creativecommons]{eptcs}

\providecommand{\event}{MEMICS 2016} 
\usepackage{breakurl}             
\usepackage{underscore}           

\usepackage[cmex10]{amsmath}
\usepackage{amssymb}
\usepackage{amsfonts}

\usepackage{amsthm}
\usepackage{enumerate}
\usepackage{hyperref}
\usepackage{float}
\usepackage{caption}
\usepackage{graphicx}
\usepackage{array}

\urldef{\mails}\path|{icharvat, smrcka, vojnar}@fit.vutbr.cz|

\newcommand{\Hades}{\textsc{Hades}}

\newcommand{\ten}{\mathtt{en}}

\usepackage{color}

\begin{document}

\title{\Hades{}: Microprocessor Hazard Analysis via\\ Formal Verification of
Parameterized Systems}

\author{Luk\'{a}\v{s} Charv\'{a}t \qquad\qquad Ale\v{s} Smr\v{c}ka \qquad\qquad Tom\'{a}\v{s}
Vojnar
\institute{Brno University of Technology, FIT,
IT4Innovations Centre of Excellence \\ Bo\v{z}et\v{e}chova 2, 612 66 Brno,
Czech Republic \\ \mails}}

\def\titlerunning{\Hades{}: Microprocessor Hazard Analysis via Formal
Verification of Parameterized Systems}

\def\authorrunning{L. Charv\'{a}t, A. Smr\v{c}ka, and T. Vojnar}

\maketitle

\begin{abstract}

\Hades{}\footnote{\url{www.fit.vutbr.cz/research/groups/verifit/tools/hades/}} is a
fully automated verification tool for pipeline-based microprocessors that aims
at flaws caused by improperly handled data hazards. It focuses on
single-pipeline microprocessors designed at the register transfer level (RTL)
and deals with read-after-write, write-after-write, and write-after-read
hazards. \Hades{} combines several techniques, including data-flow analysis,
error pattern matching, SMT solving, and abstract regular model checking. It
has been successfully tested on several microprocessors for embedded
applications.

\end{abstract}

\section{Introduction} \label{sec:introduction}

Implementation of pipeline-based execution of instructions in purpose-specific
microprocessors, often used, e.g., in embedded applications,  is an error-prone
task, which implies a~need of proper verification of the resulting designs.
Formal verification of such microprocessors---despite they are much simpler than
common processors for mainstream computing---is a very challenging task.
One way how to deal with it is to develop a~set of verification techniques
specialised in checking absence of a~certain kind of errors in such
microprocessors.
Here, the main idea is that, this way, a~high degree of automation and
scalability can be achieved since only parts of a~design related to a~specific
error are to be investigated.
The above idea has been followed, e.g., in the works~\cite{MTV12,MTV14} that
proposed fully automated approaches for (1) checking correctness of individual
execution of processor instructions and (2) for verifying absence of
read-after-write (RAW) hazards when the instructions are pipelined.
In~\cite{EUROCAST15}, the approach was extended by covering write-after-write
(WAW) and write-after-read (WAR) hazards.

To be more precise, an \emph{RAW hazard} arises when an instruction writes to a
storage that some later instruction reads, but it is possible for the later
instruction to read an old value being rewritten by the earlier instruction.
A \emph{WAW hazard} refers to a situation when an instruction writes to a
storage and rewrites a result stored by some later instruction which already
finished its execution.
A \emph{WAR hazard} arises when a later instruction write to a destination
before it is read by the previous instruction.
There are also non-data hazards.
\emph{Structural hazards} deal with sharing resources by instructions in a
pipeline.
\emph{Control hazards} arise when an instruction is executed improperly due to
an unfinished update of a program counter.
This paper, however, concentrates on data hazards only.


In particular, the paper presents the \Hades{} tool, developed by VeriFIT
research group at FIT BUT, that implements a slightly improved version of the
approaches proposed in \cite{MTV14,EUROCAST15}.
Namely, after briefly discussing related works, we specify how the input of
\Hades{} looks like, we describe its architecture and implementation, and
provide experimental results on a larger set of microprocessors than
in~\cite{MTV14,EUROCAST15}. Moreover, we include a more detailed discussion of
the needed verification time and its decomposition to the computing times
needed by the different analysis phases implemented in \Hades{}.
We close the paper by a discussion of possible future improvements of the
\Hades{} tool.

\paragraph{Related Work.}

Verifying that there are no hazards in a pipelined microprocessor is quite
crucial.
Hence, it has become a native part of checking conformance between an~RTL
design and a~formally encoded description of an instruction set architecture
(ISA), and many approaches with formal roots have been proposed for this
purpose. 
Among them, one can find, e.g., the following approaches
\cite{BurchCAV94,DillFMCAD96,AagaardCHARME03,PixleyDATE09Hector,BeyerFMCAD10,VelevICCAD11,XieDATE14}.
However, these methods typically require a~significant manual user
intervention---either in a~form of specifying the consistent state of the
microprocessor or defining predicates describing pipeline behaviour.
Compared with such approaches, \Hades{} does not aim at full conformance
checking of RTL and ISA implementations.
Instead, it addresses one specific property---namely, absence of problems
caused by pipeline hazards.
On the other hand, \Hades{} is almost fully automated---the user is required to
identify the architectural resources (such as registers and memory ports) and
the program counter only.


\section{Input Models}

\Hades{} focuses on microprocessors with a single pipeline and in-order
execution.
The tool expects storages (registers and memories) to have a~unit write and zero
read delay.
Multicycle delay storages can be easily simulated by a chain of unit storages.
The tool also assumes that pipeline internal registers which carry data
interchanged between programmer visible storages are controlled by stall and
clear signals.

\begin{figure}[t]
  \centering
  \includegraphics[width=0.9\linewidth]{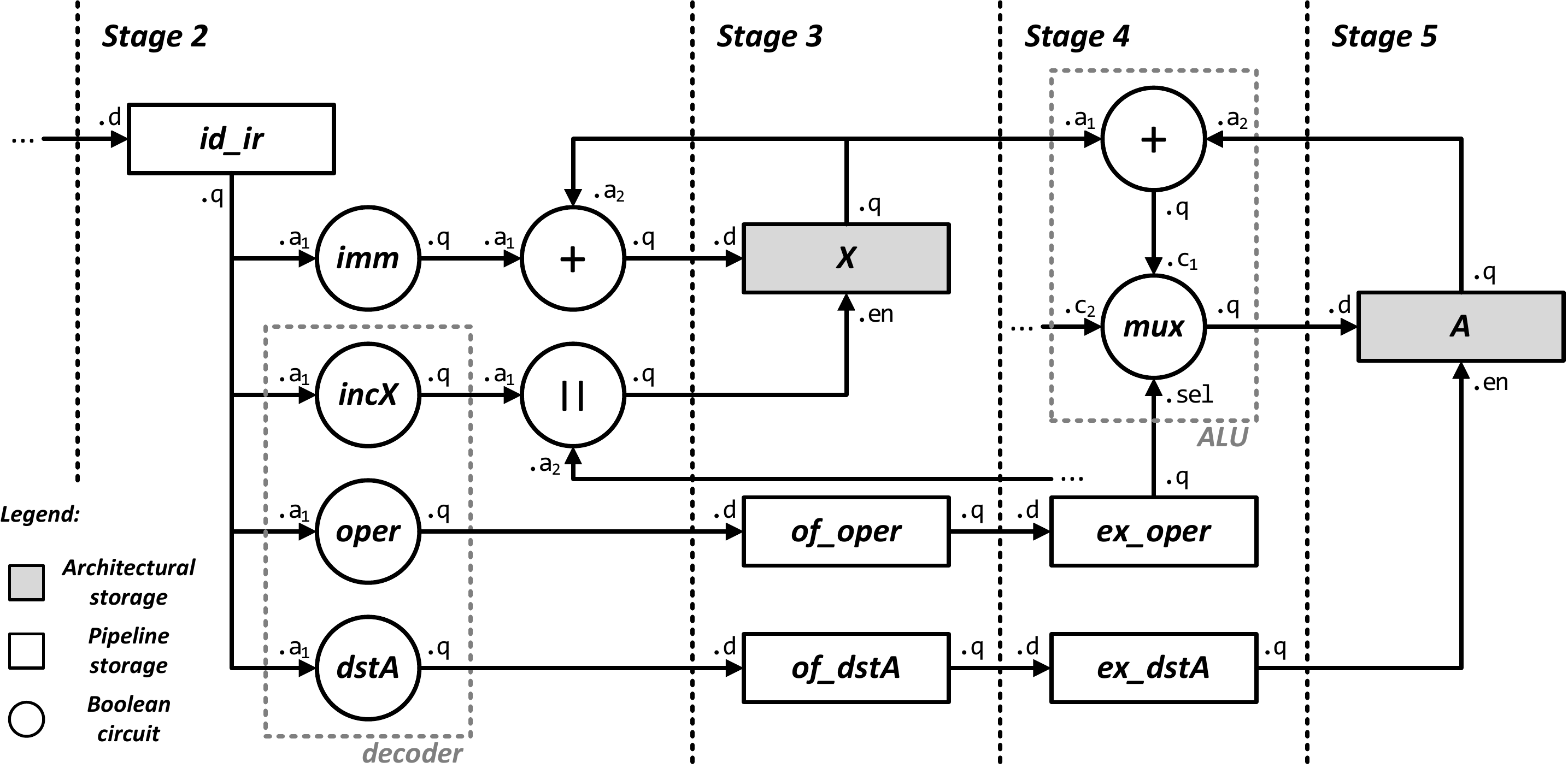}
  \caption{A processor structure graph of a~part of a~CPU with an accumulator
    architecture.}
  \label{fig:compacc}
\end{figure}

The tool expects the processor under verification to be described by a so-called
\emph{processor structure graph} (PSG in short) which represents the internal
structure of the processor.
A PSG is an oriented graph that consists of vertices (storages or boolean
circuits) and edges (control and data connections).
An example of a simple PSG is depicted in Figure~\ref{fig:compacc}.
It shows a~part of a~simple microprocessor with an accumulator architecture with
two architectural registers: $X$ (a~memory index register) and $A$ (an
accumulator).
For the sake of brevity, the PSG does not exhibit control connections of
pipeline registers.
In the CPU, an instruction fetched from the memory is stored into the storage
$\mathit{id\_ir}$ representing the instruction register.
The decoder determines the type of the operation of arithmetic logic unit and
identifies its destination by activating the appropriate enable connection
($\ten$) of the $X$ or $A$ register.
An early auto-increment of register $X$ can be performed in stage $3$.
Such a~feature allows the CPU to execute sequences of instructions working
with juxtaposed data in the memory without a~penalty (brought, e.g., by
unnecessary stalls of the pipeline) which would be present if the update of
$X$ was done in a~later stage.

A design of a processor on the register transfer level (RTL) written in a common
hardware design language like~VHDL~or~Verilog can be easily converted into a
PSG.

\section{The Verification Approach of \Hades{}}

The verification approach of \Hades{} was proposed in \cite{MTV14,EUROCAST15}. 
It leverages the current advances in SMT solvers for bit-vector logic and in
formal verification of systems with a parameterized number of processes---for
short, referred to as \emph{parameterized systems} (PSs) below.
The main idea is to reduce the problem of finding hazards that may arise when
executing an in advance unknown number of in advance unknown
instructions\footnote{Note that one cannot simply restrict the checking to a
number of instructions given by the number of pipeline stages since the
processor can get to different internal states after having processed some
number of instructions of some kind.} to a parametric verification problem where
the successive instructions are modelled by processes, which gradually pass
through the processor.
In particular, it turns out that one can use the common notion of PSs operating
on a~linear topology where the processes (i.e., instructions being executed)
may perform local transitions or universally/existentially guarded global
transitions~\cite{ClarkeVMCAI06,NamjoshiVMCAI07,HolikVMCAI13}.

More precisely, the approach consists of the following steps: (1)~a data-flow
analysis intended to distinguish particular stages of the pipeline,
(2)~a~consistency check of a~correct implementation of the particular pipeline
stages, (3)~a~static analysis identifying constraints over data-paths of
instructions that can potentially cause data hazards, (4)~generation of a~PS
modelling mutual interaction between potentially conflicting instructions, and
(5)~an analysis of the constructed parameterized system.

\paragraph{Identification of Pipeline Stages.} \label{sec:dataflow}

%
A~simple data flow analysis is used to derive the number of pipeline stages
implemented in a given processor and to assign storages and logic functions into
the pipeline stages.
A~\emph{pipeline stage} is defined as a~sub-graph of the PSG responsible for
executing a~single-cycle step of an instruction.
The pipeline stage of a~PSG vertex (representing some storage or function) is
given by the minimum number of cycles needed to propagate data from the input
of the program counter (assumed to be in a~fictive stage~$0$) to the output
of the given vertex.

\paragraph{Consistency Checking.}\label{sec:consistency}

The second step of the method is consistency checking that checks whether the
flow logic assures a~correct in-order execution of all instructions through all
the identified pipeline stages.
This step checks whether the flow logic obeys a~set of rules that express how
the control connections (i.e., enable, stall, and clear signals) of storages in
adjacent pipeline stages should be set. 
In short, the rules require that an instruction carried by a~pipeline stage
cannot be fragmented, duplicated, or lost.
In particular, a~strengthened variant of the rules proposed
in~\cite{MishraDATE02} is used.


\paragraph{Static Detection of Potential Hazards.}

Next, a~static hazard analysis over the PSG with annotated pipeline stages is
performed to identify a~finite set of so-called \emph{hazard cases}. 
Each hazard case describes one possible source of a hazard.
%
%
A~hazard case consists of a~programmer visible source storage (i.e., a~register
or a~writing port of the memory), target storage, reading and writing stages,
and an~influence path describing how data propagate between the stages.
Since the~definition of a~hazard case~speaks~about storages, their access
stages, and the path along which the problematic data is transferred, it is not
related to a~single instruction only but to an entire~class~of~instructions.

\paragraph{Generation of PSs Modelling the Possible Hazards.}

In this stage, a~PS for each identified hazard case is generated.
The main component of the PS is a finite automaton whose instances represent
instructions passing the pipeline.
A~state of the automaton identifies the class of instructions that the
particular instance represents\footnote{Three classes are distinguished---write
instructions, read instructions, and other instructions.}, the execution stage
into which the instruction got, and the conditions that must hold for the
instruction to proceed such that a flow of data along the path associated with
the given hazard case is caused.
The transitions of the automata can be guarded by referring to the states of the
automata representing instructions that surround the given instruction in the
pipeline.
Their generation is pruned by checking whether the conditions behind the states
of the involved automata do not exclude each other.
Further, regular sets of initial and bad configurations are generated.
Initial configurations represent simply an arbitrary sequence of instructions
waiting for entry into the pipeline.
Bad configurations are specified separately for the different types of hazards
considered---e.g., for RAW hazards, they say that a~later instruction finished
reading before an~earlier instruction committed writing.

\paragraph{Analysis of the Generated PS.}

As the last step, it is verified that the bad configurations are not reachable
from the initial configurations in the generated PS.
For that, \emph{abstract regular model checking} can be used \cite{VojnarCAV04}.

\section{\Hades{} Implementation}

\begin{figure}[t]
    \centering
    \includegraphics[width=0.9\linewidth]{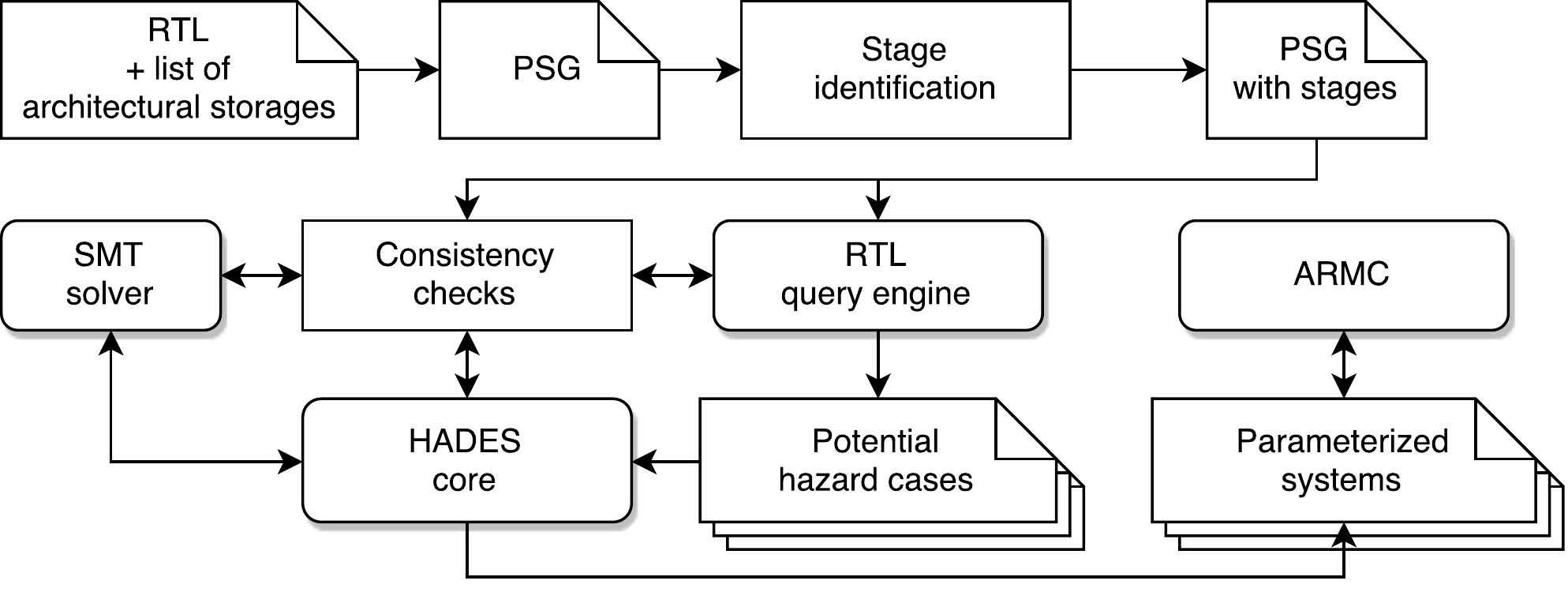}
    \caption{\Hades{} architecture.}
    \label{fig:architecture}
\end{figure}


The \Hades{} tool implements the above sketched approach and consists of several
components depicted in Figure~\ref{fig:architecture}.
\Hades{} reads in an RTL description of the processor to be verified and
converts it into its internal PSG representation. 
Currently, \Hades{} supports the RTL format of CodAl which is an architectural
description language for processor design \cite{Codal}.
For other RTL languages like VHDL and Verilog where architectural storages are
not explicitly identified, a~list of architectural storages with an explicit
identification of the program counter must be provided.

The input PSG is normalized and simplified (conditional branching is replaced by
multiplexors, value propagation is applied, redundant nodes and edges are
removed, etc.). 
For that, the \emph{RTL query engine} of \Hades{}, which allows one to search
for data-paths and substitute parts of the RTL design described by a PSG, is
used.
The engine uses a LISP-like syntax both for queries and their output, and it can
handle basic RTL constructs like signals, registers, logic gates, as well as
memory and its ports.

Subsequently, pipeline stages are identified by a simple data-flow analysis.
Intuitively, the analysis propagates so far computed stages forward through the
PSG, always taking the minimum of values incoming to a vertex and adding one
whenever a storage other than a read port (which has a zero delay) is passed.

Next, instances of the consistency rules for the particular design are derived
using RTL query engine.
The rules are checked using an SMT solver for bit-vector logic.
\Hades{} is compatible with all SMT solvers accepting SMT2 formula description.
In particular, for the below experiments, Z3~\cite{DeMoura08Z3} was used.


Further, given a PSG with annotated stages, the \Hades{} core repeatedly
utilizes the RTL query engine (written in C++) and the SMT solver to extract
potential hazard cases and to generate the appropriate PSs for them.
The generated PSs are then checked using the abstract regular model checker of
\cite{bouajjani:antichain} (implemented in OCaml over the Timbuk tree automata
library \cite{timbuk}, however, tree automata degenerated to word automata are
used only).

%
Note that the different hazard cases are are independent, and hence, in the
future, the generation of the PSs and their verification can be run in
parallel.

%

%
%


\enlargethispage{5mm}

\vspace*{-2mm}{\section{Experimental
Evaluation}\label{sec:experiments}\vspace*{-1mm}

\begin{table*}[t]
    \caption{Experimental results.}
    \label{tbl:verif-times}
    \scriptsize%
    \centering
    \begin{tabular*}{0.99\textwidth}{@{\extracolsep{\fill}} l l | c | c | c c c | c c c c | c | c }
      \hline \hline
      \textbf{Processor} / &  & Simpl.   & Data Flow    & \multicolumn{3}{c|}{Consistency} & \multicolumn{4}{c|}{Parameterized System}  & Total    & Hazard     \\
      \quad variant        &  & Time [s] & Analysis [s] &
      \multicolumn{3}{c|}{Checking [s]}& \multicolumn{4}{c|}{Generation and Verification [s]}      & Time [s] & Cases [\#] \\
      \hline                                             
                           &  &          &              & rtl     & smt   & core  & rtl   & smt   & armc  & core  &          &            \\
\hline \hline
\textbf{TinyCPU} & S        & 0.05  & 0.01 & $<$0.01  & 0.25     & 0.49       & 0.01     & 0.38     &  5.44     &  6.71      & 13.34  & 5         \\
                 & SA       & 0.06  & 0.02 & $<$0.01  & 0.33     & 0.60       & 0.02     & 1.00     & 11.58     & 20.84      & 34.45  & 8         \\
                 & B        & 0.05  & 0.01 & $<$0.01  & 0.25     & 0.44       & 0.01     & 0.38     &  5.08     &  5.95      & 12.17  & 5         \\
                 & BA       & 0.07  & 0.02 & $<$0.01  & 0.33     & 0.63       & 0.03     & 1.03     & 11.02     & 18.28      & 31.41  & 8         \\
                 & SF       & 0.06  & 0.02 & $<$0.01  & 0.30     & 0.51       & 0.02     & 0.77     & 10.82     & 13.89      & 26.39  & 11        \\
                 & SFA      & 0.07  & 0.02 & $<$0.01  & 0.34     & 0.68       & 0.04     & 1.88     & 20.42     & 43.09      & 66.54  & 18        \\
\hline                                            
\textbf{SPP8}    & S        & 0.27  & 0.04 & 0.01     & 0.43     & 0.85       & 0.05     & 2.02     & 20.81     & 36.24      & 60.72  & 27        \\
                 & B        & 0.25  & 0.03 & 0.01     & 0.40     & 0.82       & 0.07     & 2.16     & 20.35     & 43.19      & 67.28  & 27        \\
\hline                                            
\textbf{SPP16}   & S        & 0.27  & 0.05 & 0.01     & 0.44     & 0.90       & 0.04     & 1.90     & 19.99     & 36.33      & 59.93  & 27        \\
                 & B        & 0.30  & 0.05 & 0.01     & 0.43     & 0.88       & 0.07     & 2.16     & 19.75     & 42.29      & 65.94  & 27        \\
\hline                                            
\textbf{Codea2}  & SF       & 0.81  & 0.13 & 0.01     & 0.59     & 1.04       & 0.94     & 32.91    & 224.73    & 527.34     & 788.49 & 239       \\
\hline                                            
\textbf{CompAcc} & SFA      & 0.27  & 0.04 & 0.01     & 0.54     & 1.06       & 0.11     & 5.60     & 65.83     & 98.05      & 171.87 & 38        \\
                 & BFA      & 0.28  & 0.05 & 0.01     & 0.55     & 1.04       & 0.28     & 7.74     & 66.03     & 158.56     & 234.94 & 53        \\
\hline                                            
\textbf{DLX5}    & S        & 0.47  & 0.08 & 0.01     & 1.09     & 2.23       & 0.22     & 8.96     & 140.40    & 205.69     & 359.15 & 25        \\
                 & SA       & 0.54  & 0.10 & 0.01     & 1.12     & 2.44       & 0.37     & 17.54    & 250.78    & 460.75     & 733.65 & 59        \\
                 & B        & 0.62  & 0.12 & 0.01     & 1.07     & 2.40       & 0.33     & 9.47     & 138.55    & 316.08     & 468.65 & 25        \\
                 & BA       & 0.65  & 0.12 & 0.01     & 1.15     & 2.69       & 0.48     & 19.28    & 247.98    & 745.16     & 1017.52& 59        \\ 
      \hline \hline
    \end{tabular*}
    \vskip1mm
    \begin{tabular*}{0.9\textwidth}{@{\extracolsep{\fill}} l l l l l l l l}
      S & Stalling Logic & \quad B & Bypassing Logic & \quad F & Flag Register(s) & \quad A & Auto-increment Logic \\
    \end{tabular*}
\end{table*}


We have tested \Hades{} on five processors:
\emph{TinyCPU} is a~small 8-bit processor, mainly used for testing new
verification methods.
\emph{SPP8} is an~8-bit ipcore with 3 pipeline stages, 16 general-purpose
registers, and a~RISC instruction set with 9 instructions.
\emph{SPP16} is a~16-bit variant of SPP8 with a~more complex memory model.
\emph{Codea2} is a~16-bit processor for signal processing applications.
It is equipped with 16 general-purpose registers, 15 special registers, a~flag
register, and an~instruction set including 41 instructions where each may use up
to 4 available addressing modes.
\emph{CompAcc} is an~8-bit processor based on an accumulator architecture.
Finally, \emph{DLX5} is a~5-staged 32-bit processor able to execute a~subset of
the instruction set of the~DLX architecture~\cite{Patternson12Book} (with no
floating point support).

Compared with~\cite{MTV12,MTV14,EUROCAST15}, we enriched the number of
variants for the above introduced processors, which gave us 17 unique test cases
in total.
The variants of the particular processors differ in the following aspects: (i)
the way how data hazards are avoided (pipeline stalling and clearing, data
bypassing), (ii)~the presence of flag / status registers, and (iii) utilization
of so-called \emph{auto-increment} (AI) logic.
The AI logic is a feature allowing for an early incrementation\footnote{The
incrementation typically takes place in an execution stage of the processor's
pipeline.} of the value of a~register for memory addressing just before
(pre-increment) or right after (post-increment) it is read.
The AI feature usually brings a~more efficient execution of sequences of
instructions accessing the processor's memory (e.g., computation upon long
arrays in cyber-security CPUs), but it also introduces potential WAW and WAR
hazards that must be handled properly.

Besides the modifications in our test cases, we improved the \Hades{} tool
as well.
This includes an addition of dynamic programming techniques (e.g., paths found in
PSG are hashed and reused) and a~faster pipe-based communication (instead
of previously used file-based) between the \Hades{} core and the RTL query engine.

We conducted a series of experiments on a~PC with Intel Core i7-3770K @ 3.50GHz
and 16 GB RAM with results shown in Table~\ref{tbl:verif-times}.
The first columns give the verified processor, its variant, the time needed for
the~PSG simplification and its data flow analysis.
The next columns give the duration of the consistency checking and the time
spent by verification of the PSs that are created for each hazard case.
The times are split to the times consumed by the different parts of the \Hades{}
architecture.

The following column gives the overall verification time, which remains in the
order of minutes even for complex designs.
Moreover, \Hades{} also scales well with the growing size of the processor
data-path as can be seen by comparing the times obtained for \emph{SPP8} and
\emph{SPP16}.
It should be noted that the amount of time consumed by the tool's core can be
reduced by using a~direct API of the SMT solver used instead of the current
implementation that relies on exporting (potentially large) formulas in the
\texttt{smt2} file format.
(On the other hand, the current implementation does not depend on any particular
SMT solver.)
Finally, the last column represents the number of hazard cases that had to be
generated and checked.
This number differs from the one computed in~\cite{MTV14,EUROCAST15} due to
\Hades{} newly does not include hazard cases on the program counter among data
hazards.
These cases will be treated in separate control hazard detection phase, which is
currently under implementation.
Note that each hazard case represents a~separate task so the part of generation
and verification of PSs can be parallelized in the future.

During the experiments, we identified a~flaw in a~RAW hazard resolution when
accessing the data memory in a~development version of the \emph{SPP8} processor.


\vspace*{-1mm}\section{Conclusions and Future Work} \label{sec:conclusion}

%

%
%

\enlargethispage{5mm}

We have presented the main ideas, architecture, and evaluation of \Hades{}---a
tool for fully-automated discovery of data hazards in pipelined microprocessors.
In the future, we plan to extend \Hades{} with methods for verification of other
processor features, such as control hazards.
We also plan to parallize some parts of \Hades{} and extend it with a compiler
from VHDL and Verilog IP~cores to the \textsc{Hades} input format.


\paragraph*{Acknowledgement.} The work was supported by the Czech Science
Foundation project 14-11384S, the IT4IXS: IT4Innovations Excellence in Science
project (LQ1602), and the internal BUT project FIT-S-14-2486.


\bibliography{./hwverification}
\bibliographystyle{eptcs}

\end{document}